\documentclass[twocolumn]{autart}    

\usepackage{cite}
\usepackage{subcaption}
\usepackage{graphicx}
\usepackage{amsmath,amsfonts,mathrsfs}
\usepackage{hyperref}
\usepackage{float}
\usepackage{stfloats}
\usepackage[nocomma]{optidef}
\usepackage{booktabs}
\usepackage{xcolor}
\usepackage{makecell}
\usepackage{algorithm}
\usepackage{algpseudocode}
\usepackage{enumerate}

\newtheorem{assumption}{Assumption}
\newtheorem{remark}{Remark}
\newtheorem{lemma}{Lemma}
\newtheorem{theorem}{Theorem}
\newtheorem{corollary}{Corollary}

\begin{document}

\begin{frontmatter}
\title{Robust Data-Driven Predictive Control for Unknown Linear Time-Invariant Systems\thanksref{footnoteinfo}}

\thanks[footnoteinfo]{The work was supported by the National Natural Science Foundation of China through Project No. 62173287 and the Research Grants Council of the Hong Kong Special Administrative Region under the Early Career Scheme through Project No. 27206021.}

\author{Kaijian Hu}\ead{kjhu@eee.hku.hk},    
\author[Corresponding]{Tao Liu}\ead{taoliu@eee.hku.hk} 

\address{Department of Electrical and Electronic Engineering, The University of Hong Kong, Hong Kong SAR, and the HKU Shenzhen Institute of Research and Innovation, Shenzhen, China.}
\corauth[Corresponding]{Corresponding author}


\begin{abstract}
This paper presents a new robust data-driven predictive control scheme for unknown linear time-invariant systems by using input-state-output or input-output data based on whether the state is measurable. To remove the need for the persistently exciting (PE) condition of a sufficiently high order on pre-collected data, a set containing all systems capable of generating such data is constructed. Then, at each time step, an upper bound of a given objective function is derived for all systems in the set, and a feedback controller is designed to minimize this bound. The optimal control gain at each time step is determined by solving a set of linear matrix inequalities. We prove that if the synthesis problem is feasible at the initial time step, it remains feasible for all future time steps. Unlike current data-driven predictive control schemes based on behavioral system theory, our approach requires less stringent conditions for the pre-collected data, facilitating easier implementation. Further, the proposed predictive control scheme features an infinite prediction horizon, potentially resulting in superior overall control performance compared to existing methods with finite prediction horizons. The effectiveness of our proposed methods is demonstrated through application to an unknown and unstable batch reactor.
\end{abstract}

\end{frontmatter}

\section{Introduction}

Model-based control greatly depends on precise mathematical models, which can be challenging to obtain in practice \cite{BisoffiAndrea2020Dsou}. To tackle this issue, data-driven control (DDC), in particular for linear time-invariant (LTI) systems, has been introduced in the literature and gained considerable attention from different research fields \cite{BruntonStevenL2019DSaE}.

In recent years, many DDC methods have been developed for LTI systems, such as iterative learning control \cite{ChenYangquan1999Ilc}, model-free adaptive control \cite{HouZhongsheng2014MFAC}, reinforcement learning control \cite{Lopez2023}, and data-driven predictive control \cite{BerberichJulian2021DMPC, CoulsonJeremy2018DPCI}. Most of these existing DDC methods need to pre-collect adequate and informative data. For instance, in order to identify an unknown LTI system, the pre-collected data must satisfy the persistently exciting (PE) condition of a sufficiently high order\cite{TangiralaArunK2015PoSI}. Similarly, to ensure the parameter estimation convergence in model-free adaptive control, the pre-collected input data must satisfy the PE condition\cite{HouZhongsheng2014MFAC}. Furthermore, all DDC methods based on the Willems' Fundamental Lemma \cite{WillemsJanC2005Anop} mandate the pre-collected data to meet the PE condition (see \cite{DePersisClaudio2020FfDC, Hu2023, BerberichJulian2021DMPC, LuppiAlessandro2022Odso, DePersisClaudio2021LloL} for some examples). However, collecting a set of data from an unknown system that satisfies the PE condition can be challenging in practice, especially in the case of unstable systems \cite{Talebi2022}. 

To relax or remove the requirement of the PE condition, several approaches have been proposed recently\cite{vanWaardeHenkJ2020WFLf, vanWaardeHenkJ2020Fndt, Waarde2023, SteentjesTomRV2022Odci}. The concept of collective persistency of excitation is introduced in \cite{vanWaardeHenkJ2020WFLf} to remove the requirement on the PE condition for a single data sequence, but the PE condition on the combination of multiple sequences is still required. The matrix versions of S-lemma and Finsler's lemma are introduced to remove PE conditions in \cite{vanWaardeHenkJ2020Fndt} and  \cite{Waarde2023}, respectively, and a robust DDC framework is built to achieve the control targets of all systems that are compatible with the pre-collected data without the PE condition. Since the systems compatible with the given data contain the true system, the designed controller can also achieve the control target for the true system. This framework has been applied to solve various control problems, such as state feedback control\cite{vanWaardeHenkJ2020Fndt, Waarde2023} and $H_\infty$ control\cite{SteentjesTomRV2022Odci, hu2022}. 

On the other hand, model predictive control (MPC) is a crucial and efficient control technique for systems with constraints and has wide applications in industrial processes\cite{QIN2003733, MAYNE2000789, CamachoE.F2004Mpc}. By solving an optimization problem at every time step, MPC calculates optimal control input sequences, with only the first control input in the sequence being implemented in the system. As each subsequent time step occurs, the optimization problem is reevaluated using updated system measurements \cite{RakoviSasaV2018HoMP}. Similar to other model-based control approaches, MPC depends on an accurate model for predicting the system behavior during the optimization process.

However, for a system whose model is unknown or has unknown uncertainties, predicting its future behavior is challenging. To address this issue, the authors in \cite{KothareMayureshV1996Rcmp} propose a robust MPC scheme. This method establishes an upper bound for a given objective function within the optimization problem, followed by the design of a state feedback controller aimed at minimizing this upper bound.  It is worth noting that information on a nominal model of the system is still required for implementing this method. 

Recently, data-driven predictive control, which does not rely on the explicit model, is developed in \cite{BerberichJulian2021DMPC, BerberichJulian2021LtMf, CoulsonJeremy2018DPCI}. As discussed before, these methods also require the PE condition of a sufficiently high order on the pre-collected data; otherwise, they may not function properly, as the predicted input-output trajectory might not be compatible with the true system. How to remove the requirement on the PE condition for data-driven predictive control for unknown LTI systems subject to constraints remains open and deserves attention.

To address the aforementioned issues, this paper develops a novel, robust, data-driven predictive control scheme for unknown LTI systems, eliminating the need for the PE condition on pre-collected data. First, inspired by \cite{Waarde2023}, we construct a data-represented set containing all systems capable of generating such data. Then, motivated by \cite{KothareMayureshV1996Rcmp}, rather than solving a min-max MPC problem for all systems in the set, which is computationally intractable under the traditional MPC framework, we aim to find an upper bound of the objective function for all systems and design a feedback controller to minimize this bound at each time step. The optimal control gain is determined by solving a set of linear matrix inequalities (LMIs) at each time step. The feasibility of the synthesis problem is ensured by the feasibility of the problem at the initial time step. We begin by addressing the unconstrained data-driven predictive control problem, assuming the state of the unknown LTI system is measurable. Then, we consider cases with both input and output constraints. Finally, we extend the proposed method to scenarios where the state is not measurable by only using input-output data.

The structure of this paper is as follows. Section \ref{sc-problemfor} gives the unknown LTI system to be studied and some necessary preliminaries. In Section \ref{sc-disturbace}, a new robust data-driven predictive control problem is proposed. By assuming the system state is measurable, Section \ref{sc-mainresults} develops a robust data-driven predictive control scheme for the unconstrained LTI system by using pre-collected input-sate-output data. Section \ref{sc-constraint} discusses the robust data-driven control design for the system with input and output constraints also by using pre-collected input-sate-output data. Section \ref{sc-arx} extends the method proposed in Section \ref{sc-constraint} to the case where the system state is unmeasurable by using pre-collected input-output data.
In Section \ref{sc-numericalEx}, the effectiveness of the proposed method is demonstrated by using a batch reactor. Finally, Section \ref{sc-conclusion} concludes the paper.

\textbf{Notation:} Let $\mathbb{R}$, $\mathbb{Z}$, and $\mathbb{N}$ denote the set of real numbers, integers, and natural numbers, respectively. $I_{n}$ is the $n\times n$ identity matrix; $0_{n\times m}$ is the $n\times m$ zero matrix; the subscript $n$ and $n\times m$ may be omitted when the dimensions are clear from the context. For a matrix $M\in\mathbb{R}^{n\times m}$, $M^\top$ stands for its transpose;  $M^{-1}$ denotes its inverse if it is square and non-singular; $M^\dagger$ denotes its pseudo-inverse; $\text{ker}(M)$ denotes the kernel of $M$.  For a symmetric $M$, $M<0\ (M\leq 0)$ denotes that $M$ is negative (semi-)definite. $\text{diag}$($M_1,\dots, M_s$) represents a block diagonal matrix consisting of matrices $M_i\in\mathbb{R}^{n_i\times m_i}$, $i=1,\dots,s$. The notation $\|\cdot\|_2$ represents the Euclidean norm of a vector and the induced norm of a matrix, respectively. Given a signal ${z}(k):\mathbb{Z}\to \mathbb{R}^n$, define ${z}_{[k,k+T]}=[{z}(k), {z}(k+1),\dots,{z}(k+T)]$, $k\in\mathbb{Z}$ and $T\in\mathbb{N}$.

\section{Problem Formulation}\label{sc-problemfor}
Consider a multiple-input and multiple-output (MIMO) LTI system
\begin{subequations}\label{eq-system}
\begin{align}
x(k+1) &= A x(k)+Bu(k),\label{eq-systema}\\
y(k)&=C x(k)+D u(k),\label{eq-systemb}
\end{align}
\end{subequations}
where $u(k)\in\mathbb{R}^{m}$, $x(k)\in\mathbb{R}^{n}$, and $y(k)\in\mathbb{R}^p$ are the system input, state, and output, respectively; $A\in\mathbb{R}^{n\times n}$, $B\in\mathbb{R}^{n\times m}$, $C\in\mathbb{R}^{p\times n}$, and $D\in\mathbb{R}^{p\times m}$ are the system matrices.

Without loss of generality, we make the following assumption for the system (\ref{eq-system}), which is widely utilized in the literature of data-driven control \cite{DePersisClaudio2020FfDC}.
\begin{assumption}\label{as-as1}
    The system matrices $A$, $B$, $C$, and $D$ are unknown, but their dimensions are known, i.e., the numbers $m$, $n$, and $p$ are known.
\end{assumption}

If the system matrices $A$, $B$, $C$, and $D$ are known, then the model predictive control problem at  each time $k$ can be formulated as the following receding horizon optimization problem \cite{CamachoE.F2004Mpc}
\begin{mini!}|s|[1]
{\substack{u(k+i|k),\\i=0,\dots, L-1}}{J(k)}{\label{pr-arx1}}{}
\addConstraint{x(k+1+i|k)&=A x(k+i|k)+B u(k+i|k),}{\label{pr-arx2}}
\addConstraint{y(k+i|k)&= C x(k+i|k)+D u(k+i|k),}{\label{pr-arx3}}
\end{mini!}
where the objective function $J(k)$ is defined as
\begin{align}
    J(k) = \sum_{i=0}^{L-1}\begin{Vmatrix}
        Q^{\frac{1}{2}}y(k+i|k)
    \end{Vmatrix}_2^2+\begin{Vmatrix}
        R^{\frac{1}{2}}u(k+i|k)
    \end{Vmatrix}_2^2;\label{eq-obj}
\end{align}
$L$ is the prediction horizon; $Q\in\mathbb{R}^{p\times p}$ and $R\in\mathbb{R}^{m\times m}$ are two positive definite weight matrices; $x(k|k)=x(k)$ and $y(k|k)=y(k)$ are the measured state and output of the system \eqref{eq-system} at time $k$, respectively; $x(k+i|k)$ and $y(k+i|k)$, $\forall i=1,\dots, L-1$ are the predicted state and output based on the predictive model \eqref{pr-arx2} and \eqref{pr-arx3}, respectively; $u(k+i|k)$, $\forall i=0,\dots, L-1$ are the control inputs to be calculated. At time $k$, only the first step input $u(k)=u(k|k)$ obtained by (\ref{pr-arx1}) is implemented to the system (\ref{eq-system}), and this process is repeated at the next time $k+1$.

Since the system matrices in (\ref{eq-system}) are unknown, directly solving the optimization problem (\ref{pr-arx1}) is impossible. To solve this issue, we can use the pre-collected input-state-output/input-output data to identify the model of (\ref{eq-system}) and then solve the problem (\ref{pr-arx1}). Alternatively, we can use pre-collected data to design a data-driven predictive controller \cite{BerberichJulian2021DMPC, CoulsonJeremy2018DPCI} based on Willems' fundamental lemma\cite{WillemsJanC2005Anop}. 

All the abovementioned methods require that the pre-collected input data satisfy the PE condition with a sufficiently high order, which is essential for uniquely identifying either the system model \cite{TangiralaArunK2015PoSI} or behavior \cite{WillemsJanC2005Anop}. However, collecting a set of data from an unknown system that satisfies the PE condition can be challenging, especially in the case of an unknown unstable system\cite{Talebi2022}.  If a given set of pre-collected data from the system (\ref{eq-system}) does not fulfill the PE condition, it may be generated by other systems also in the form of (\ref{eq-system}) but with different system matrices. Consequently, the accurate model of the system cannot be retrieved, or the system behavior cannot be determined uniquely based on the pre-collected data, which makes the MPC problem \eqref{pr-arx1} or the data-driven predictive control problem \cite{BerberichJulian2021DMPC, BerberichJulian2021LtMf} unsolvable.

Before ending this section, the Schur complement lemma and matrix Finsler's lemma are recapped, which will be used later.
\begin{lemma}[\cite{1994Lmii}]\label{le-lemma1}
    Let $Q\in \mathbb{R}^{n\times n}$, $P\in \mathbb{R}^{m\times m}$ be symmetric matrices and $R\in \mathbb{R}^{n\times m}$ be an $n\times m$ matrix. The following three statements are equivalent
    \begin{enumerate}[(i)]
        \item $\begin{bmatrix}
            Q & R\\
            R^\top & P
            \end{bmatrix} < 0$,
        \item $Q<0$ and $P-R^\top Q^{-1} R < 0$,
        \item $P<0$ and $Q-RP^{-1}R^\top < 0$.
    \end{enumerate}
\end{lemma}
\begin{lemma}[\cite{Waarde2023}]\label{le-slemma}
Consider two symmetric matrices $M,N\in \mathbb{R}^{(q+s)\times (q+s)}$. Let $M$ and $N$ be partitioned as
\begin{align}
M=
\begin{bmatrix}
M_{11} & M_{12}\\
M_{12}^\top & M_{22}
\end{bmatrix}, N=
    \begin{bmatrix}
    N_{11} & N_{12}\\
    N_{12}^\top & N_{22}
    \end{bmatrix}\label{eq-partition}
\end{align}
with $M_{11}, N_{11}\in \mathbb{R}^{q\times q}$, $M_{22}, N_{22}\in \mathbb{R}^{s\times s}$, and $M_{12}, N_{12}\in \mathbb{R}^{q\times s}$. Suppose that $M_{22}\geq 0$, $N_{22}\geq 0$, $N_{11}-N_{12}N_{22}^\dagger N_{12}^\top=0$, and $\text{ker}(N_{22})\subseteq \text{ker}(N_{12})$. Then, the following two statements are equivalent

\begin{enumerate}[(i)]
    \item $[I_q, Z] M [I_q, Z]^\top<0$ for any $Z\in\mathbb{R}^{q\times s}$ satisfying $[I_q, Z] N [I_q, Z]^\top = 0$,
    \item There exist scalars $\alpha\geq 0$ and $\beta > 0$ such that 
    \begin{align}
        M-\alpha N\leq \begin{bmatrix}
            -\beta I_q & 0\\
            0 & 0_{s\times s}
        \end{bmatrix}.\nonumber
    \end{align}
\end{enumerate}
\end{lemma}


\section{Robust Data-Driven Predictive Control Problem}\label{sc-disturbace}
To remove the PE condition on the pre-collected input data, this paper presents a robust data-driven predictive control scheme for the unknown LTI system (\ref{eq-system}) by using input-state-output when the system state is measurable. We further extend the proposed method to the case where the system state is unmeasurable by only relying on input-output data. First, we will consider the case that the state is measurable.

Let (${u}_{[0,T-1]}$, ${x}_{[0,T]}$, $y_{[0,T-1]}$) be a pre-collected input-state-output trajectory of the system (\ref{eq-system}), and define the following related matrices
\begin{subequations}
\begin{align}
U &= [{u}(0), {u}(1), \dots, {u}(T-1)],\label{eq-Ui}\\
Y &= [{y}(0), {y}(1), \dots, {y}(T-1)],\\
X &= [{x}(0), {x}(1), \dots, {x}(T-1)],\\
X_{+}&= [{x}(1), {x}(2), \dots, {x}(T)].\label{eq-Xp}
\end{align}
\end{subequations}

The data matrices (\ref{eq-Ui})-(\ref{eq-Xp}) satisfy
\begin{subequations}\label{eq-ehwdes1}
\begin{align}
X_{+}&=AX+B U,\label{eq-dataa}\\
Y &= CX+DU.\label{eq-datab}
\end{align}
\end{subequations}
Here, we abuse the notations a bit and still use $A\in\mathbb{R}^{n\times n}$, $B\in\mathbb{R}^{n\times m}$, $C\in\mathbb{R}^{p\times n}$, and $D\in\mathbb{R}^{p\times m}$ to denote all the possible matrices that satisfy \eqref{eq-dataa} and \eqref{eq-datab}. Further, let $\Sigma$ be the set comprising all of these possible matrices, i.e.,
\begin{align}
    \Sigma=\{(A,B,C,D)|(\ref{eq-dataa})\text{ and }(\ref{eq-datab})\}.\label{eq-set1}
\end{align}

Since the actual system matrices of the system \eqref{eq-system}  belong to the set $\Sigma$, a robust data-driven predictive controller designed for all systems within $\Sigma$ will also be effective for the true system \eqref{eq-system}. 

To facilitate the robust data-driven predictive controller design, we integrate the two positive definite weight matrices $Q$ and $R$ of the objective function $J(k)$ into the description of $\Sigma$. Firstly, we left-multiply both sides of (\ref{eq-datab}) by $Q^{\frac{1}{2}}$ and have 
\begin{align}
    Q^{\frac{1}{2}}Y = Q^{\frac{1}{2}}CX+Q^{\frac{1}{2}}DU.\label{eq-databQ}
\end{align}
Secondly, we introduce the following identity equation into the set $\Sigma$
\begin{align}
    R^{\frac{1}{2}}U = 0_{m\times n}X+R^{\frac{1}{2}}U.\label{eq-dataaux}
\end{align}
Since (\ref{eq-databQ}) is equivalent to (\ref{eq-datab}), and (\ref{eq-dataaux}) is an identity equation, the set $\Sigma$ can be equivalently described by
\begin{align}
    \Sigma=\{(A,B,C,D)|(\ref{eq-dataa})\text{ and }(\ref{eq-databQ})\text{ and }(\ref{eq-dataaux})\}.\label{eq-set1e}
\end{align}

Further, (\ref{eq-dataa}), (\ref{eq-databQ})\text{, and }(\ref{eq-dataaux}) together can be rewritten in a matrix form as
\begin{align}
[I_{n+m+p}, Z]H=0\label{eq-ehwdes}
\end{align}
by defining
\begin{align}
Z&=\begin{bmatrix}
A & B\\
Q^{\frac{1}{2}}C & Q^{\frac{1}{2}}D\\
0_{m\times n} & R^{\frac{1}{2}}
\end{bmatrix},\label{eq-Zdef}\\
H&=[X_{+}^\top, (Q^{\frac{1}{2}}Y)^\top, (R^{\frac{1}{2}}U)^\top, -X^\top, -U^\top]^\top.\label{eq-Hdef}
\end{align}
Right-multiplying both sides of (\ref{eq-ehwdes}) by the transpose of $[I_{n+m+p}, Z]H$ and defining $N=H {H}^\top$ yield the following quadratic matrix equation
\begin{align}
[I_{n+m+p}, Z] N [I_{n+m+p}, Z]^\top = 0.\label{eq-condition}
\end{align}
Utilizing the fact that $\Phi\Phi^\top=0$ is equivalent to $\Phi=0$ for any real matrix $\Phi$, we have (\ref{eq-ehwdes}) is equivalent to (\ref{eq-condition}). Therefore, the set $\Sigma$ can be redefined as 
\begin{align}
    \Sigma=\{(A, B, C, D)|(\ref{eq-condition})\}.\label{eq-SigmaFinal}
\end{align}
As will be discussed in Remark \ref{rmrk2}, the quadratic matrix equation \eqref{eq-condition} will be used in the robust data-driven predictive controller design to characterize the set $\Sigma$, which contains all possible systems capable of generating the pre-collected data  (\ref{eq-Ui})-(\ref{eq-Xp}).

Since the pre-collected data  (\ref{eq-Ui})-(\ref{eq-Xp}) cannot uniquely determine the system model or behavior, the MPC problem \eqref{pr-arx1} is not solvable.  As an alternative, a min-max receding horizon optimization problem can be formulated as
\begin{mini!}|s|[1]
{\substack{u(k+i|k), \\ i=0,\dots, L-1}} {\max_{(A,B,C,D)\in \Sigma} J(k)}{\label{pr-data}}{}
\addConstraint{x(k+1+i|k)&=A x(k+i|k)+B u(k+i|k),}{\label{pr-datasys2}}
\addConstraint{y(k+i|k)&= C x(k+i|k)+D u(k+i|k).}{\label{pr-datasys3}}
\end{mini!}
The min-max problem, as described, aims to minimize the worst-case value of the objective function $J(k)$ among all systems in $\Sigma$. This problem is known to be computationally intractable under the traditional MPC framework, as mentioned in \cite{KothareMayureshV1996Rcmp}. 

To address these computational challenges, we propose a novel, robust, data-driven predictive control scheme for the unknown system \eqref{eq-system}. This new control scheme can be seen as a data-driven version of the predictive control scheme developed in \cite{KothareMayureshV1996Rcmp}. By leveraging data-driven techniques, the proposed control scheme aims to overcome the limitations of the traditional MPC framework and provide a more efficient and robust solution.


\section{Unconstrained Predictive Control Using Input-State-Output Data}\label{sc-mainresults}
In this section, we propose a new robust data-driven predictive control scheme for the unknown LTI system (\ref{eq-system}) by using pre-collected input-state-output data. Since the min-max problem \eqref{pr-data} is not computationally tractable, we turn to solve an alternative problem inspired by \cite{KothareMayureshV1996Rcmp}. The alternative problem involves finding an upper bound $\eta(k)\in\mathbb{R}$ of  the objective $J(k)$ across all systems in $\Sigma$ and designing a state feedback controller 
\begin{equation}\label{controller}
u(k+i|k)=F_kx(k+i|k)
\end{equation}
with the control gain $F_k=F(k)$ that minimizes the upper bound $\eta(k)$ at each time $k$.  Obviously, a feasible solution to the problem at each time $k$ will provide a robust data-driven predictive control scheme for the unknown system \eqref{eq-system}. 

The results of the derived upper bound and the designed controller are presented in the following theorem.

\begin{theorem}\label{th-theorem1}
Consider the system (\ref{eq-system}) under Assumption \ref{as-as1}. Given a set of pre-collected input-state-output data (${u}_{[0,T-1]}$, ${x}_{[0,T]}$, $y_{[0,T-1]}$) of (\ref{eq-system}), if there exist scalars $\alpha_k=\alpha(k)\geq 0$, $\beta_k=\beta(k)>0$, $\eta_k=\eta(k) > 0$, a matrix $S_k=S(k)\in\mathbb{R}^{m\times {n}}$, and a positive definite matrix $\Gamma_k=\Gamma(k)\in\mathbb{R}^{{n}\times{n}}$ at time $k$ such that the following optimization problem (\ref{pr-dpc}) is feasible, 
\begin{mini!}|s|
{\substack{\alpha_k,\beta_k, \eta_k,\\ S_k,\Gamma_k}}{\eta_k\label{eq-objective}}{\label{pr-dpc}}{}
\addConstraint{\begin{bmatrix}
-\eta_k & x(k|k)^\top\\
x(k|k) & -\Gamma_k
\end{bmatrix}\leq 0,}\label{eq-gammax}{}
\addConstraint{\mathcal{M}_k-\alpha_k\mathcal{N}\leq \begin{bmatrix}
    -\beta_k I_{n+m+p} & 0\\
    0 & 0
\end{bmatrix} ,}\label{eq-dVwN}{}
\end{mini!}
 then $x(k|k)^\top \Gamma_k^{-1} x(k|k)$ is an upper bound of the objective function $J(k)$ for all systems in $\Sigma$, where $\mathcal{M}_k$ and $\mathcal{N}$ in \eqref{eq-dVwN} are defined as
\begin{align}
\mathcal{M}_k&=\begin{bmatrix}
    -\Gamma_k& 0 & 0 & 0 & 0\\
    0 & -I_{m+p} & 0 & 0 & 0\\
    0 & 0 &\Gamma_k & S_k^\top & 0\\
    0 & 0 &S_k & 0 & S_k\\
    0 & 0 &0 & S_k^\top & -\Gamma_k
\end{bmatrix},\\
\mathcal{N}&=
\begin{bmatrix}
    N & 0 \\
    0 & 0_{n\times n}
\end{bmatrix},\label{mat_N}
\end{align}
and $N$ in \eqref{mat_N} is defined right before (\ref{eq-condition}). Further, the control gain matrix $F_k$ in the state feedback controller ${u}(k+i|k)=F_k x(k+i|k)$, $i=0,\dots, L-1$ that  minimizes the upper bound is $F_k=S_k\Gamma_k^{-1}$. 
\end{theorem}
\begin{pf} Under the controller \eqref{controller}, the closed-loop system of (\ref{pr-datasys2})-(\ref{pr-datasys3}) is 
\begin{subequations}
\begin{align}
x(k+i+1|k)=\bar{A}_k x(k+i|k)\label{eq-clsystema}\\
y(k+i|k)=\bar{C}_k x(k+i|k)\label{eq-clsystemb}
\end{align}
\end{subequations} 
for all $(A, B, C, D)\in\Sigma$ and $i=0,\dots, L-1$ where $\bar{A}_k=\bar A(k)=A+BF_k$ and $\bar{C}_k=\bar C(k)=C+DF_k$. Define $P_k=\Gamma_k^{-1}$ and the quadratic function
\begin{align}
V({x}(k+i|k))={x}(k+i|k)^\top P_k {x}(k+i|k).\nonumber
\end{align}
We claim that if the LMI (\ref{eq-dVwN}) is feasible, then $V({x}(k|k))$ is an upper bound of $J(k)$ for all systems in $\Sigma$.

Since $-\Gamma_k<0$, applying Lemma \ref{le-lemma1} to (\ref{eq-dVwN}) gives
\begin{align}
M_k-\alpha_k N\leq \begin{bmatrix}
    -\beta_k I_{n+m+p} & 0\\
    0 & 0
\end{bmatrix}, \label{eq-MtauN}
\end{align}
where $M_k$ is defined as
\begin{align}
M_k=\text{diag}\bigg(\begin{bmatrix} -\Gamma_k& 0 \\ 0 & -I_{m+p}\end{bmatrix}, \begin{bmatrix}\Gamma_k & S_k^\top \\ S_k & S_k\Gamma_k^{-1} S_k^\top \end{bmatrix}\bigg).\label{eq-Mdef}
\end{align}
Substituting the control gain matrix $F_k=S_k\Gamma_k^{-1}$ into (\ref{eq-Mdef}) gives
\begin{align}
M_k=\text{diag}\bigg(\!\begin{bmatrix} -\Gamma_k& 0 \\ 0 & -I_{m+p}\end{bmatrix}\!, \begin{bmatrix}\Gamma_k & \Gamma_k F_k^\top\\ F_k\Gamma_k & F_k\Gamma_k F_k^\top\end{bmatrix}\!\bigg).\label{eq-MK}
\end{align}
By partitioning $M_k$ into the form of (\ref{eq-partition}), we have
\begin{align}
M_{22}=\begin{bmatrix}
    \Gamma_k & \Gamma_k F_k^\top\\
    F_k\Gamma_k & F_k\Gamma_k F_k^\top
    \end{bmatrix}=\begin{bmatrix}
        I_n\\
        F_k
    \end{bmatrix}\Gamma_k \begin{bmatrix}
        I_n\\
        F_k
    \end{bmatrix}^\top,\nonumber
\end{align}
based on which we get $M_{22}\geq 0$. Furthermore, partitioning $N$ into the form of (\ref{eq-partition}) gives
\begin{align}
    N = \begin{bmatrix}
        N_{11} & N_{12}\\
        N_{12}^\top & N_{22}
        \end{bmatrix}=\begin{bmatrix}
        H_1 H_1^\top & H_1 H_2^\top\\
        H_2 H_1^\top & H_2 H_2^\top
    \end{bmatrix},\nonumber
\end{align}
where $H_1=[X_{+}^\top, (Q^{\frac{1}{2}}Y)^\top, (R^{\frac{1}{2}}U)^\top]^\top, H_2=[-X^\top,$ $-U^\top]^\top$. By calculation, we have $N_{22}\geq 0$ and $N_{11}-N_{12}N_{22}^\dagger N_{12}^\top=0$. Since $N_{12}=-Z N_{22}$, we can obtain $\text{ker}(N_{22})\subseteq \text{ker}(N_{12})$. By applying Lemma \ref{le-slemma}, we derive that (\ref{eq-MtauN}) holds if and only if 
\begin{align}
    [I_{n+m+p}, Z] M_k [I_{n+m+p}, Z]^\top< 0 \label{eq-Mcon}
\end{align}
holds for any $Z$ satisfying $[I_{n+m+p}, Z] N [I_{n+m+p}, Z]^\top\!\!\!=\!0$.\!

Based on the definition of the set $\Sigma$ in \eqref{eq-SigmaFinal}, inequality (\ref{eq-MtauN}) holds if and only if inequality (\ref{eq-Mcon}) holds for all systems in $\Sigma$. With \eqref{eq-clsystema} and \eqref{eq-clsystemb}, inequality (\ref{eq-Mcon}) can be rewritten as
\begin{align}
    [I_{n+m+p}, \bar{Z}] \bar{M}_k [I_{n+m+p}, \bar{Z}]^{\top} < 0,\label{eq-Mbar}
\end{align}
with $\bar{M}_k=\text{diag}(-\Gamma_k,-I_{m+p},\Gamma_k)$ and $\bar{Z}= [\bar{A}_k^\top, $ $(Q^{\frac{1}{2}}\bar{C}_k)^\top, $ $(R^{\frac{1}{2}}F_k)^\top]^\top$. Substituting the definition of $\bar{M}_k$ into (\ref{eq-Mbar}) gives
\begin{align}
\left[\begin{array}{c}
    \bar{A}_k \\
    Q^{\frac{1}{2}}\bar{C}_k\\
    R^{\frac{1}{2}}F_k
\end{array}\right]\Gamma_k \left[\begin{array}{c}
    \bar{A}_k \\
    Q^{\frac{1}{2}}\bar{C}_k\\
    R^{\frac{1}{2}}F_k
\end{array}\right]^\top\!\!\!- \left[\begin{array}{ccc}
    \Gamma_k & 0 & 0\\
    0 & I_p & 0\\
    0 & 0 & I_m
    \end{array}\right] < 0.\label{eq-phipsidxx}
\end{align}
Applying Lemma \ref{le-lemma1} to (\ref{eq-phipsidxx}) twice gives
\begin{align}
\bar{A}_k^\top P_k \bar{A}_k-P_k+ \bar{C}_k^\top Q \bar{C}_k+F_k^\top R F_k< 0,\nonumber
\end{align}
which is equivalent to
\begin{align}
&{x}(k+i|k)^\top\big(\bar{A}_k^\top P_k \bar{A}_k-P_k+\bar{C}_k^\top Q \bar{C}_k\nonumber\\
&+F_k^\top R F_k\big){x}(k+i|k)< 0, \forall {x}(k+i|k)\neq 0.\label{eq-phipsid}
\end{align}
On the other hand, the difference between $V({x}(k+i+1|k))$ and $V({x}(k+i|k))$ is 
\begin{align}
\Delta V({x}(k+i|k)) =& {x}(k+i+1|k)^\top P_k {x}(k+i+1|k)\!\nonumber\\
&\!-{x}(k+i|k)^\top P_k {x}(k+i|k).\label{eq-dV}
\end{align}
Substituting the constraint (\ref{eq-clsystema}) into (\ref{eq-dV}) gives  
\begin{align}
    \!\Delta V({x}(k\!+\!i|k))\!=\!{x}(k\!+\!i|k)^\top\!(\bar{A}_k^\top P_k \bar{A}_k\!-\!P_k){x}(k\!+\!i|k).\!\!\label{eq-deltaV}
\end{align}
Substituting (\ref{eq-deltaV}) into (\ref{eq-phipsid}) gives
\begin{align}
&\Delta V(x(k+i|k))+{x}(k+i|k)^\top\big(\bar{C}_k^\top Q \bar{C}_k\nonumber\\
&+F_k^\top R F_k\big){x}(k+i|k)< 0.\label{eq-deltaVp}
\end{align}
Substituting (\ref{eq-clsystemb}) and the state feedback controller $u(k+i|k)=F_k x(k+i|k)$ into (\ref{eq-deltaVp}) gives
\begin{align}
&\Delta V({x}(k+i|k))+\big(y(k+i|k)^\top Q y(k+i|k)\nonumber\\
&+u(k+i|k)^\top R u(k+i|k)\big)< 0.\label{eq-fin}
\end{align}
Summing (\ref{eq-fin}) from $i=0$ to $i=L-1$ gives
\begin{align}
&-V({x}(k|k))+J(k)+V({x}(k+L|k))< 0.\!\label{eq-finx1}
\end{align}
Since $V({x}(k+L|k))\geq 0$, we have
\begin{align}
J(k) < V({x}(k|k)). \label{eq-maxJ}
\end{align}
Therefore,  if (\ref{eq-dVwN}) is feasible, $V({x}(k|k))=x(k|k)^\top \Gamma_k^{-1}$ $x(k|k)$ is an upper bound of the objective function $J(k)$ for all systems in $\Sigma$. 

Further, applying Lemma \ref{le-lemma1} to (\ref{eq-gammax}) gives 
\begin{align}
    V({x}(k|k))=x(k|k)^\top \Gamma_k^{-1} x(k|k)\leq \eta_k. \label{eq-VGamma}
\end{align}
In addition, the problem (\ref{pr-dpc}) is convex as the objective function (\ref{eq-objective}) is convex and both the constraints (\ref{eq-gammax}) and (\ref{eq-dVwN}) are LMIs. Therefore, if the problem (\ref{pr-dpc}) is feasible, then its globally optimal solutions exist and minimize the objective function  $\eta_k$.  Based on (\ref{eq-VGamma}), the upper bound $V({x}(k|k))$ on $J(k)$ is minimized under the feedback controller ${u}(k+i|k)=F_k x(k+i|k)$, $i=0,\dots, L-1$ with $F_k=S_k\Gamma_k^{-1}$.
\end{pf}

\begin{remark}\label{rmrk1}
It is easy to verify that  (\ref{eq-finx1}) also holds for the infinite prediction horizon case. As a result, $V({x}(k|k))$ is an upper bound on the objective function $J(k)$ regardless of the prediction horizon length $L$. Consequently, we can minimize the upper bound of $J(k)$ even with an infinite prediction horizon ($L=\infty$) by solving the optimization problem (\ref{pr-dpc}).
\end{remark}

\begin{remark}\label{rmrk2}
To prove $V({x}(k|k))$ is an upper bound of $J(k)$, i.e., to ensure (\ref{eq-maxJ}) holds, it is crucial to make the QMI (\ref{eq-Mcon}) contain the information of the two weight matrices $Q$ and $R$ in $J(k)$. Given that the matrix $M_k$  in (\ref{eq-Mcon}) is fixed (see the definition of $M_k$ in \eqref{eq-MK}), the only way to do so is to incorporate $Q$ and $R$ into the matrix $Z$.  Moreover, in order to get the equivalent conditions of (\ref{eq-MtauN}), i.e., (\ref{eq-Mcon}) under \eqref{eq-condition}, by applying Lemma \ref{le-slemma}, the quadratic matrix terms in (\ref{eq-condition}) and (\ref{eq-Mcon}) must have the same matrix $Z$, this is guaranteed by introducing (\ref{eq-databQ}) and (\ref{eq-dataaux}) into (\ref{eq-set1e}), which is the reason that we incorporate $Q$ and $R$ to characterize the set $\Sigma$ in \eqref{eq-SigmaFinal}.
\end{remark}

\begin{remark}\label{re-data}
    In Theorem \ref{th-theorem1}, two types of data are employed. One is the pre-collected data, i.e., $U$, $Y$, $X$, and $X_+$. These data sequences are utilized in (\ref{eq-dVwN}) and remain unchanged throughout the control process. The other one is the online measurement data, i.e., $x(k)$, which is utilized in (\ref{eq-gammax}) and needs to be updated at each time $k$.
\end{remark}

\begin{algorithm}
    \caption{Robust Data-Driven Predictive Control Scheme Using Input-State-Output Data}\label{al-dpc}
    \begin{algorithmic}[1]
        \State Collect the input-state-output sequence;
        \State Set $k=0$;
        \Loop
            \State Let $x(k|k)=x(k)$ and solve (\ref{pr-dpc});
            \State Set $F_k=S_k\Gamma_k^{-1}$;
            \State Apply the input $u(k)=F_kx(k)$;
            \State Set $k=k+1$;
        \EndLoop
    \end{algorithmic}
\end{algorithm}

Based on Theorem  \ref{th-theorem1}, the robust data-driven predictive control scheme is developed by solving the optimization problem (\ref{pr-dpc}) in a receding horizon fashion, as shown in Algorithm \ref{al-dpc}. However, two important issues remain open: 1) whether the optimization problem (\ref{pr-dpc}) is feasible at each time $k$ with $k>0$ if it is feasible at $k=0$; 2) whether the designed controller stabilizes the system (\ref{eq-system}). The answers to these issues are presented in the following theorem.

\begin{theorem}\label{th-fea}
Consider the system (\ref{eq-system}) under Assumption \ref{as-as1}. Given a set of pre-collected input-state-output data (${u}_{[0,T-1]}$, ${x}_{[0,T]}$, $y_{[0,T-1]}$) of (\ref{eq-system}), if the problem (\ref{pr-dpc}) is feasible at the initial time $k=0$, then
\begin{enumerate}[(i)]
    \item the problem (\ref{pr-dpc}) is feasible at any time $k \in \mathbb{N}$,
    \item the system (\ref{eq-system}) is stabilized by the designed controller $u(k)=F_kx(k)$ with $F_k=S_k\Gamma_k^{-1}$, $\forall k \in \mathbb{N}$.
\end{enumerate}
\end{theorem}
\begin{pf}
To prove the statement (i), let $\alpha_k^*$, $\beta_k^*$, $\eta_k^*$, $S_k^*$ and $\Gamma_k^*$ be the optimal solution of the problem (\ref{pr-dpc}) at time $k$. Define $P_k^*={\Gamma_k^*}^{-1}$. Without loss of generality, we assume that the initial state $x(0)\neq 0$. The feasibility of (\ref{pr-dpc}) at $k=0$ implies that (\ref{eq-gammax}) and (\ref{eq-dVwN}) hold at $k=0$. From (\ref{eq-gammax}), we get
\begin{align}
{x}(0|0)^\top P_0^* {x}(0|0) \leq \eta_0^*.\label{eq-ppxx}
\end{align}
From (\ref{eq-fin}), we get $\Delta V(x(i|0)) < 0$,  i.e.,
\begin{align}
{x}(i\!+\!1|0)^\top P_0^* {x}(i\!+\!1|0)\!-\!{x}(i|0)^\top P_0^* {x}(i|0)\!<\!0.\! \label{eq-ppx}
\end{align}
Combining (\ref{eq-ppxx}) with (\ref{eq-ppx}) gives
\begin{align}
&{x}(i+1|0)^\top P_0^* {x}(i+1|0)<{x}(i|0)^\top P_0^* {x}(i|0) \nonumber\\
<&\dots<{x}(0|0)^\top P_0^* {x}(0|0)\leq \eta_0^*.\label{eq-ppxx1}
\end{align}
By setting $i=1$ in (\ref{eq-ppxx1}), we have 
\begin{align}
{x}(1|0)^\top P_0^* {x}(1|0)<{x}(0|0)^\top P_0^* {x}(0|0)\leq \eta_0^*\label{eq-ppxx4}
\end{align}
for all systems in $\Sigma$.
Further, at $k=1$, we have the sampled state $x(1)=x(1|1)$, which can also be considered as the one step prediction $x_t(1|0)$ of the predictive model $(A_t, B_t, C_t, D_t)$ from the initial condition $x(0|0)$, i.e.,  $x(1|1)=x_t(1|0)=(A_t+B_t F_0^*)x(0|0)$ with $F_0^*=S_0^*{\Gamma_0^*}^{-1}$. Here, to distinguish from the other system matrices in $\Sigma$, we use $(A_t, B_t, C_t, D_t)$ to denote the real system matrices of  \eqref{eq-system}. Since $(A_t, B_t, C_t, D_t)\in\Sigma$, from \eqref{eq-ppxx4} we have 
\begin{equation}
{x_t}(1|0)^\top P_0^* {x_t}(1|0)<{x}(0|0)^\top P_0^* {x}(0|0) \leq \eta_0^*, 
\end{equation}
which is equivalent to 
\begin{align}
{x}(1|1)^\top P_0^* {x}(1|1) <{x}(0|0)^\top P_0^* {x}(0|0)\leq \eta_0^*.\label{eq-ppxx3}
\end{align}
Thus, (\ref{eq-gammax}) is feasible at time $k=1$ by choosing $\Gamma_1=\Gamma_0^*$ and $\eta_1=\eta_0^*$.

If we consider  (\ref{eq-dVwN}) alone, it only depends on the pre-collected data, and thus its feasible solution space of (\ref{eq-dVwN}) remains unchanged for any time $k\geq0$. Therefore, there exists at least one solution $\alpha_1=\alpha_0^*$, $\beta_1=\beta_0^*$, $S_1=S_0^*$ and $\Gamma_1=\Gamma_0^*$ satisfying (\ref{eq-dVwN}) at time $k=1$.

Since both the constraints (\ref{eq-gammax}) and (\ref{eq-dVwN}) in  (\ref{pr-dpc}) are feasible at $k=1$, the problem (\ref{pr-dpc}) is feasible at time $k=1$. By repeating the above process, we claim that (\ref{pr-dpc}) is feasible for any $k>0$ if it is feasible at $k=0$. This completes the proof of statement (i).

To prove the statement (ii), we choose $V^*({x}(k|k))=x(k|k)^{\top}P^*_kx(k|k)$ as the Lyapunov function candidate. Since $\Gamma_1^*={P_1^*}^{-1}$ is the optimal solution of the problem (\ref{pr-dpc}) at $k=1$, we have
\begin{align}
    {x}(1|1)^\top P_1^*{x}(1|1)\leq {x}(1|1)^\top P_0^*{x}(1|1).\label{eq-p1p}
\end{align} 
Combining (\ref{eq-ppxx3}) with (\ref{eq-p1p}) gives
\begin{align}
    {x}(1|1)^\top P_1^*{x}(1|1)< {x}(0|0)^\top P_0^*{x}(0|0).\label{eq-p1px}
\end{align} 
By repeating the process from \eqref{eq-ppxx} to \eqref{eq-p1px}, we have 
\begin{align}
    {x}(k+1|k+1)^\top P_{k+1}^*{x}(k+1|k+1)< {x}(k|k)^\top P_k^*{x}(k|k)\label{eq-p1pxa}
\end{align} 
hold for $k=1,2,\dots$, which yields
\begin{align}
    \Delta V^*({x}(k|k))=&{x}(k+1|k+1)^\top P_{k+1}^*{x}(k+1|k+1)\nonumber\\
    &-{x}(k|k)^\top P_k^*{x}(k|k)\nonumber\\
    < &0.
\end{align}
Since $V^*({x}(k|k)) > 0$ and $\Delta V^*({x}(k|k))< 0$ for any $x(k)=x(k|k)\neq 0$, the system (\ref{eq-system}) is stabilized by the designed controller. Together with the proof of statement (i), this completes the proof of the theorem.
\end{pf}
\begin{remark}\label{re-stability}
Based on the proof of Theorem \ref{th-fea}, if the problem (\ref{pr-dpc}) is feasible at the initial time $k=0$, then the designed controller stabilizes all systems in $\Sigma$. Consequently, all systems in $\Sigma$, including the real system, must be stabilizable; otherwise, the controller may not exist. 
\end{remark}

\begin{remark}\label{re-compareDMPC}
To design a data-driven predictive control scheme based on behavioral system theory, most of the existing results (e.g., \cite{BerberichJulian2021DMPC, CoulsonJeremy2018DPCI}) usually require two basic assumptions: i) the pre-collected input data is PE of a sufficiently high order and ii) the unknown system \eqref{eq-system} is controllable. In our paper, we remove the first assumption and relax the second assumption on controllability to stabilizability. Further, we do not need to check whether the unknown system is stabilizable in advance but just solve the problem (\ref{pr-dpc}) directly, as the feasibility of the problem (\ref{pr-dpc}) at $k=0$ is a sufficient condition of the stabilizability of all the systems in $\Sigma$. 
\end{remark}
    
\begin{remark}
Compared with the robust model predictive control scheme proposed in \cite{KothareMayureshV1996Rcmp}, which requires knowledge of the nominal model, our method solely relies on pre-collected data and does not need any explicit model information.  This makes the proposed method more advantageous and practical for situations where model information is not available or difficult to obtain.
\end{remark}

\begin{remark}
Similar to Section IV in \cite{KothareMayureshV1996Rcmp}, our control method can be applied to tracking problems when the reference input $u_r$, state $x_r$, and output $y_r$ are known. In such cases, the proposed robust data-driven predictive control scheme can be utilized to stabilize the error system
\begin{subequations}
    \begin{align}
    \Delta x(k+1) &= A \Delta x(k)+B\Delta u(k),\label{eq-systema_err}\\
    \Delta y(k)&=C \Delta x(k)+D\Delta u(k),\label{eq-systemb_err}
    \end{align}
\end{subequations}
with $\Delta x(k)=x(k)-x_r$, $\Delta u(k)=u(k)-u_r$, and $\Delta y(k)=y(k)-y_r$, which is equivalent to track the desired output $y_r$ by driving the system to the reference input $u_r$ and state $x_r$. 
\end{remark}

\section{Constrained Predictive Control Using Input-State-Output Data}\label{sc-constraint}
In the previous section, we propose a robust data-driven predictive control scheme for the unknown system (\ref{eq-system}) without taking any constraints into account. However, in practical systems, constraints on the input and output,  such as saturation constraints\cite{Essick2014}, linear inequality constraints\cite{Bolognani2015}, and norm constraints\cite{He2021}, are often present\cite{1994Lmii, KothareMayureshV1996Rcmp}. 

In this section, we extend the proposed control scheme to include input and output constraints. Due to space limitations, only norm constraints are considered, but the treatment of the other types of constraints can be done in a similar way. 

Consider the following input and output norm constraints
\begin{align}
\|u(k)\|_2\leq u_{\max},\ \|y(k)\|_2\leq y_{\max}, \label{eq-constraints}
\end{align} 
where $u_{\max}$ and $y_{\max}$ are two positive constants. The main idea here is to formulate the input and output constraints as LMIs and then incorporate them into the data-driven predictive control scheme developed in Section \ref{sc-mainresults}. 

However, incorporating the output constraints requires knowledge of the unknown matrices $C$ and $D$, which cannot be determined uniquely, as the pre-collected data does not satisfy the PE condition. To overcome this issue,  we construct a set $\Sigma_{C,D}$ that contains all possible matrices $C$ and $D$ by using the pre-collected input-state-output data in a similar way to the construction of $\Sigma$. 

Define $Z_y=[C, D]$. By employing a similar process as in (\ref{eq-ehwdes1})-(\ref{eq-condition}), we get
\begin{align}
    [I_{p}, Z_y] N_y [I_{p}, Z_y]^\top = 0,\label{eq-ZyZy}
\end{align}
where $N_y=H_y H_y^\top$ with $H_y=[Y^\top, -X^\top, $ $-U^\top]^\top$. Consequently, the set $\Sigma_{C,D}$ that contains all possible matrices $C$ and $D$ can be expressed as 
\begin{align}
    \Sigma_{C,D} = \{(C,D)| (\ref{eq-ZyZy})\}.\label{eq-SigmaCD}
\end{align}
Next, we will design a controller to guarantee that the predicted inputs and outputs satisfy the input constraint and output constraint in \eqref{eq-constraints}, respectively.

Before presenting the main results of this section, we introduce an equivalent optimization problem to (\ref{pr-dpc}). Let $\bar{S}_k=\eta_k S_k$, $\bar{\Gamma}_k=\eta_k \Gamma_k$, $\bar{\alpha}_k=\eta_k\alpha_k$, and $\bar{\beta}_k=\eta_k\beta_k$. The problem (\ref{pr-dpc}) can be rewritten as
    \begin{mini!}|s|
        {\substack{\bar{\alpha}_k, \bar{\beta}_k, \eta_k, \\ \bar{S}_k, \bar{\Gamma}_k}}{\eta_k}{\label{op-obcons}}{}
        \addConstraint{\begin{bmatrix}
        -1 & x(k|k)^\top\\
        x(k|k) & -\bar{\Gamma}_k
        \end{bmatrix}\leq 0,}\label{eq-gammaxcon}{}
        \addConstraint{\bar{\mathcal{M}}_k-\bar{\alpha}_k\mathcal{N}\leq \begin{bmatrix}
            -\bar{\beta}_k I_{n+m+p} & 0\\
            0 & 0
        \end{bmatrix},}\label{eq-dVwNcon}{}
    \end{mini!}
    with
    \begin{align}
        \bar{\mathcal{M}}_k=\begin{bmatrix}
            -\bar{\Gamma}_k& 0 & 0 & 0 & 0\\
             0 & -\eta_k I_{m+p}& 0 & 0 & 0\\
             0 & 0 &\bar{\Gamma}_k & \bar{S}_k^\top & 0\\
             0 & 0 &\bar{S}_k & 0 & \bar{S}_k\\
             0 & 0 &0 & \bar{S}_k^\top & -\bar{\Gamma}_k
            \end{bmatrix}.\label{eq-mathcalMk}
\end{align}

The following theorem presents the results on the upper bound of $J(k)$ and the corresponding controller for the unknown system \eqref{eq-system} with the input and output constraints \eqref{eq-constraints}.

\begin{theorem}\label{th-th3}
Consider the system (\ref{eq-system}) subject to the input and output constraints (\ref{eq-constraints}). Suppose that Assumption \ref{as-as1} is satisfied. Given a set of pre-collected input-state-output data (${u}_{[0,T-1]}$, ${x}_{[0,T]}$, $y_{[0,T-1]}$) of (\ref{eq-system}), if there exist non-negative scalars $\bar{\alpha}_k$ and $\tau_k$, positive scalars $\eta_k$,  $\bar{\beta}_k$ and $\kappa_k$, a matrix $\bar{S}_k\in\mathbb{R}^{m\times {n}}$, and a positive definite matrix $\bar{\Gamma}_k\in\mathbb{R}^{{n}\times{n}}$ at time $k$ such that the following optimization problem (\ref{pr-dpc_con}) is feasible,
\begin{mini!}|s|
    {\substack{\bar{\alpha}_k, \bar{\beta}_k, \eta_k, \tau_k, \\  \kappa_k, \bar{S}_k, \bar{\Gamma}_k}} {\eta_k\label{eq-obj_con}}{\label{pr-dpc_con}}{}
    \addConstraint{(\ref{eq-gammaxcon}), (\ref{eq-dVwNcon}),}\label{eq-cons_equ}{}
    \addConstraint{\begin{bmatrix}
        -u_{\max}^2I_{m} & \bar{S}_k\\
        \bar{S}_k^\top & -\bar{\Gamma}_k
    \end{bmatrix}\leq 0,}\label{eq-inputconsrx}{}
    \addConstraint{\mathcal{M}_{y,k}-\tau_k\mathcal{N}_y\leq \begin{bmatrix}
        -\kappa_k I_{p} & 0\\
        0 & 0
    \end{bmatrix},}\label{eq-outputconsrx}{}  
\end{mini!}
then $x(k|k)^\top \eta_k \bar{\Gamma}_k^{-1} x(k|k)$ is an upper bound of the objective function $J(k)$ for all systems in $\Sigma$, where $\mathcal{M}_{y,k}$ and $\mathcal{N}_y$ in \eqref{eq-outputconsrx} are defined as
\begin{align}
    \mathcal{M}_{y,k}&=
    \begin{bmatrix}
        -y_{\max}^2 I_{p} & 0 & 0 & 0\\
        0 & \bar{\Gamma}_k        & \bar{S}_k^\top   & 0\\
        0 & \bar{S}_k             & 0            & \bar{S}_k\\
        0 & 0               & \bar{S}_k^\top   & -\bar{\Gamma}_k
    \end{bmatrix},\label{eq-mathcalMky}\\
    \mathcal{N}_y&=
    \begin{bmatrix}
        N_y & 0\\
        0 & 0_{n\times n}
    \end{bmatrix},\label{eq-Ny_e}
\end{align}
and $N_y$ in (\ref{eq-Ny_e}) is defined right after (\ref{eq-ZyZy}). Further, the state feedback control gain matrix $F_k=\bar{S}_k\bar{\Gamma}_k^{-1}$ minimizes the upper bound. Additionally, the predicted inputs and outputs satisfy the following constraints
\begin{align}
    \|u(k+i|k)\|_2&\leq u_{\max},\label{eq-inputcons}\\
    \|y(k+i|k)\|_2&\leq y_{\max}\label{eq-outputcons} 
\end{align}
for all $(C, D)$ in $\Sigma_{C,D}$ and $i=0,\dots, L-1$.
\end{theorem}
\begin{pf}
We will first show that with the feasibility of LMIs in (\ref{eq-cons_equ}), i.e., \eqref{eq-gammaxcon} and \eqref{eq-dVwNcon}, the input constraint (\ref{eq-inputcons}) and output constraint (\ref{eq-outputcons})  are satisfied if LMIs (\ref{eq-inputconsrx})  and (\ref{eq-outputconsrx}) are feasible, respectively.

To prove the input constraint part, i.e., the input constraint (\ref{eq-inputcons}) is satisfied if LMIs (\ref{eq-inputconsrx}) and (\ref{eq-cons_equ}) are feasible,  applying Lemma \ref{le-lemma1} to (\ref{eq-inputconsrx}) yields
    \begin{align}
        \bar{S}_k \bar{\Gamma}_k^{-1} \bar{S}_k^\top - u_{\max}^2 I_{m}\leq 0,
    \end{align}
which together with $\bar{\Gamma}_k>0$ implies
    \begin{align}
 \|\bar{S}_k\bar{\Gamma}_k^{-\frac{1}{2}}\|_2=\sqrt{\sigma_{\max}(\bar{S}_k \bar{\Gamma}_k^{-1} \bar{S}_k^\top)}\leq u_{\max}, \label{eq-x-umaxeta}
    \end{align}
where $\sigma_{\max}(\cdot)$ denotes the largest eigenvalue of a square matrix. 
 
On the other hand, define $P_k=\eta_k\bar{\Gamma}_k^{-1}$ and the quadratic function
    \begin{align}
    V({x}(k+i|k))={x}(k+i|k)^\top P_k {x}(k+i|k).\nonumber
    \end{align}
Since (\ref{eq-cons_equ}), i.e., \eqref{eq-gammaxcon} and \eqref{eq-dVwNcon} are feasible, we have (\ref{eq-deltaVp}) by using the same proof as that in Theorem \ref{th-theorem1}. Further, 
from (\ref{eq-deltaVp}) with $Q>0$ and $R>0$, we have 
    \begin{align}
       \Delta V({x}(k+i|k)) <0, \forall i\geq 0,
    \end{align}
    which, together with $\Delta V({x}(k+i|k))$ defined in (\ref{eq-dV}) gives
        \begin{equation}\label{eq-x-dVi}
        \begin{split}
       &x(k+i+1|k)^\top\eta_k\bar{\Gamma}_k^{-1}x(k+i+1|k)\\
       <&x(k+i|k)^\top\!\eta_k\bar{\Gamma}_k^{-1}x(k+i|k).
\end{split}
\end{equation}
    Combining (\ref{eq-VGamma}) with (\ref{eq-x-dVi}) gives
    \begin{align}
        x(k+i|k)^\top\eta_k \bar{\Gamma}^{-1}x(k+i|k) \leq \eta_k, \forall i\geq 0.\label{eq-x-GammaEta}
    \end{align}
    Since $\bar{\Gamma}_k>0$ and $\eta_k>0$, \eqref{eq-x-GammaEta} is equivalent to
    \begin{align}
        \|\bar{\Gamma}_k^{-\frac{1}{2}}x(k+i|k)\|_2\leq 1, \forall i\geq 0. \label{eq-x-invariant}
    \end{align}
Combining (\ref{eq-x-umaxeta}) with (\ref{eq-x-invariant}) gives
    \begin{align}\label{ineq_ex}
        \|\bar{S}_k\bar{\Gamma}_k^{-\frac{1}{2}}\|_2 \|\bar{\Gamma}_k^{-\frac{1}{2}}x(k+i|k)\|_2\leq u_{\max}, \forall i\geq 0.
    \end{align}
 Combining \eqref{ineq_ex} with $u(k+i|k)=\bar{S}_k\bar{\Gamma}_k^{-1}x(k+i|k)$ gives
    \begin{align}
    \|u(k+i|k)\|_2=\|\bar{S}_k\bar{\Gamma}_k^{-1}x(k+i|k)\|_2 \leq u_{\max}, \forall i\geq 0, \label{eq-x-inputconsr}
    \end{align}
i.e., the input constraint is satisfied if LMIs (\ref{eq-inputconsrx}) and (\ref{eq-cons_equ}) are feasible.

To prove the output constraint part, i.e., the output constraint (\ref{eq-outputcons}) is satisfied if LMIs (\ref{eq-outputconsrx}) and (\ref{eq-cons_equ}) are feasible, following the similar procedure from (\ref{eq-MtauN}) to (\ref{eq-Mcon}) yields that the LMI (\ref{eq-outputconsrx}) holds if and only if
\begin{align}
    [I_p,Z_y]
    M_{y,k}
    [I_p,Z_y]^\top< 0 \label{eq-x-My}
\end{align}
holds for any $Z_y$ satisfying $[I_p,Z_y] N_y [I_p,Z_y]^\top = 0$, where
\begin{align}
    M_{y,k}=\text{diag}(-y_{\max}^2 I_p, \begin{bmatrix}
        \bar{\Gamma}_k & \bar{\Gamma}_k F_k^\top\\
        F_k\bar{\Gamma}_k & F_k\bar{\Gamma}_k F_k^\top
    \end{bmatrix}),\nonumber
\end{align}
which is equivalent to 
\begin{align}
    M_{y,k}=\text{diag}(-y_{\max}^2 I_p, \begin{bmatrix}
        I_n\\
        F_k
    \end{bmatrix}\bar{\Gamma}_k \begin{bmatrix}
        I_n\\
        F_k
    \end{bmatrix}^\top).\label{eq-M_yk_sim}
\end{align}

Substituting \eqref{eq-M_yk_sim} into \eqref{eq-x-My} gives
\begin{align}
    -y_{\max}^2 I_p+Z_y\begin{bmatrix}
        I_n\\
        F_k
    \end{bmatrix}\bar{\Gamma}_k \begin{bmatrix}
        I_n\\
        F_k
    \end{bmatrix}^\top Z_y^\top< 0. \label{eq-x-My_sim}
\end{align}

Based on the definitions of $Z_y$ and $\bar C_k$ (the latter is defined right after \eqref{eq-clsystemb}), we have
\begin{align}
    \bar{C}_k=Z_y[I_n, F_k^\top]^\top.\label{eq-x-Nycons}
\end{align}
Substituting (\ref{eq-x-Nycons}) into (\ref{eq-x-My_sim}) gives
$$\bar{C}_k \bar{\Gamma}_k \bar{C}_k^\top < y_{\max}^2 I_p,$$
which implies
\begin{align}
    \|\bar{C}_k\bar{\Gamma}_k^{\frac{1}{2}}\|_2=\sqrt{\sigma_{\max}(\bar{C}_k \bar{\Gamma}_k \bar{C}_k^\top)}< y_{\max}.\label{eq-CkGammak}
\end{align}
Since (\ref{eq-cons_equ}), i.e., \eqref{eq-gammaxcon} and \eqref{eq-dVwNcon} are feasible, we have  (\ref{eq-x-invariant}) hold. Combining (\ref{eq-CkGammak}) with (\ref{eq-x-invariant}) yields
    \begin{align}
        \|\bar{C}_k x(k+i|k)\|_2 \leq y_{\max}, \forall i\geq 0.\label{eq-x-outputconsr}
    \end{align}
 Substituting \eqref{eq-clsystemb} into \eqref{eq-x-outputconsr} gives,
 $$\|y(k+i|k)\|_2\leq y_{\max}, \forall i\geq 0,$$
i.e., the output constraint (\ref{eq-outputcons}) is satisfied if LMIs \eqref{eq-outputconsrx} and (\ref{eq-cons_equ}) are satisfied. 

The rest of the poof is the same as that for Theorem \ref{th-theorem1} and thus is omitted.
\end{pf}

Similar to Theorem \ref{th-fea},  the following theorem addresses the two key questions: 1) whether the optimization problem (\ref{pr-dpc_con}) is feasible at each time $k$ with $k>0$ if it is feasible at $k=0$; 2) whether the designed controller stabilizes the unknown system (\ref{eq-system}) with the input and output constraints \eqref{eq-constraints}.

\begin{theorem}\label{th-fea_con}
    Consider the system (\ref{eq-system}) subject to the input and output constraints (\ref{eq-constraints}). Suppose that Assumption \ref{as-as1} is satisfied. Given a set of pre-collected input-state-output data (${u}_{[0,T-1]}$, ${x}_{[0,T]}$, $y_{[0,T-1]}$) of (\ref{eq-system}), if the problem (\ref{pr-dpc_con}) is feasible at the initial time $k=0$, then
    \begin{enumerate}[(i)]
        \item the problem (\ref{pr-dpc_con}) is feasible at any time $k \in \mathbb{N}$,
        \item the system (\ref{eq-system}) is stabilized by the designed controller $u(k)=F_kx(k)$ with $F_k=\bar{S}_k\bar{\Gamma}_k^{-1}$, $\forall k \in \mathbb{N}$.
    \end{enumerate}
\end{theorem}
\begin{pf}
The proof is similar to that of Theorem \ref{th-fea}, and thus is omitted.
\end{pf}

 Based on Theorem \ref{th-th3} and Theorem \ref{th-fea_con}, the proposed robust data-driven predictive control scheme for the unknown system (\ref{eq-system}) with input and output constraints is summarized in Algorithm \ref{al-dpc_con}.

 \begin{algorithm}
    \caption{Robust Data-Driven Predictive Control Scheme Using Input-State-Output Data}\label{al-dpc_con}
    \begin{algorithmic}[1]
        \State Collect the input-state-output sequence;
        \State Set $k=0$;
        \Loop
            \State Let $x(k|k)=x(k)$ and solve (\ref{pr-dpc_con});\label{eq-solvedpr}
            \State Set $F_k=\bar{S}_k\bar{\Gamma}_k^{-1}$;
            \State Apply the input $u(k)=F_kx(k)$;
            \State Set $k=k+1$;
        \EndLoop
    \end{algorithmic}
\end{algorithm}

\begin{remark}
 Comparing Algorithm \ref{al-dpc} with Algorithm \ref{al-dpc_con}, the main difference lies in the problem to be solved in step \ref{eq-solvedpr}. The former solves the problem \eqref{pr-dpc} that does not consider any constraints, while the latter solves the problem \eqref{pr-dpc_con} that takes into account input and output constraints.
\end{remark}

\begin{remark}
Compared to the problem \eqref{pr-dpc}, the problem \eqref{pr-dpc_con} adds two new LMIs, i.e., \eqref{eq-inputconsrx} and \eqref{eq-outputconsrx}, which describe input and output constraints. It is worth noting that neither the pre-collected data nor the online measurement data are used in \eqref{eq-inputconsrx}. Additionally, only the pre-collected data, i.e., $U$, $Y$, and $X$, is used in \eqref{eq-outputconsrx}. 
\end{remark}

\section{Predictive Control Scheme Using Input-Output Data}\label{sc-arx}
Motivated by the fact that system states are not always measurable in practical applications, this section extends the proposed data-driven control method by only using input-output data. To achieve this goal, we first construct a state-space representation by using input-output data.  Then, based on this data-based representation, we design the robust predictive controller for (\ref{eq-system}) by applying the results obtained in Section \ref{sc-mainresults} and Section \ref{sc-constraint}.

For simplicity, this section focuses on a specific class of the system (\ref{eq-system}) with $D=0$. Similar to \cite{DePersisClaudio2020FfDC}, we can construct a new state using the past $n$-step input and output data as
\begin{align}
    \hat{x}(k) =[&{u}(k-n)^\top, \dots, {u}(k-1)^\top, \nonumber\\
    &y(k-n)^\top, \dots, y(k-1)^\top]^\top.\label{eq-rdef}
\end{align}
Using $\hat{x}(k)$, we can establish a new state-space model that has the same input-output relationship as the original system (\ref{eq-system})\cite{GoodwinGrahamC2009AFPa, PhanMP1996Rbsa, DePersisClaudio2020FfDC}, which is described as follows: 
\begin{subequations}\label{eq-statespace}
\begin{align}
\hat{x}(k+1)&=\hat{A} \hat{x}(k)+\hat{B} {u}(k),\label{eq-statespace1}\\
y(k)&=\hat{C}\hat{x}(k),\label{eq-statespace2}
\end{align}
\end{subequations}
where $\hat{A}\in\mathbb{R}^{\hat{n}\times\hat{n}}$, $\hat{B}\in\mathbb{R}^{\hat{n}\times m}$, and $\hat{C}\in\mathbb{R}^{p\times\hat{n}}$ are the system matrices with $\hat{n}=n(m+p)$. One approach to deriving the system \eqref{eq-statespace} from \eqref{eq-system} involves three steps: i) constructing an observable subsystem that has the same input-output behavior as the original system \eqref{eq-system}\cite{GoodwinGrahamC2009AFPa}; ii) constructing an auto-regressive model with exogenous input (ARX model) using the observable subsystem \cite{PhanMP1996Rbsa}; iii) constructing the system \eqref{eq-statespace} using the ARX model \cite{DePersisClaudio2020FfDC}. It is seen that the system matrices $\hat{A}$, $\hat{B}$, and $\hat{C}$ can be represented using the information contained in the original system matrices $A$, $B$, and $C$. However, their specific forms do not influence the analysis and thus are omitted to save space.

 According to Assumption \ref{as-as1}, the system matrices $A$, $B$, and $C$ are unknown, but their dimensions are known.  Consequently, the new system matrices $\hat{A}$, $\hat{B}$, and $\hat{C}$ are unknown, while their dimensions can be calculated based on available information. Therefore, the system (\ref{eq-statespace}) satisfies Assumption \ref{as-as1} as well, and the results derived in Theorems \ref{th-theorem1}-\ref{th-fea_con} can be applied to the controller design for the system (\ref{eq-statespace}).

Let (${u}_{[0,T-1]}$, $y_{[0,T-1]}$) be a pre-collected input-output trajectory of the system (\ref{eq-statespace}). Define the following data matrices
\begin{subequations}
    \begin{align}
        \tilde{U} &= [{u}(n), {u}(n+1), \dots, {u}(T-1)],\label{eq-Uix}\\
        \tilde{Y} &= [{y}(n), {y}(n+1), \dots, {y}(T-1)],\\
        \hat{X} &= [\hat{x}(n), \hat{x}(n+1), \dots, \hat{x}(T-1)],\\
        \hat{X}_{+}&= [\hat{x}(n+1), \hat{x}(n+2), \dots, \hat{x}(T)].\label{eq-Xpx}
    \end{align}
\end{subequations}

Similar to the definitions of $Z$ in \eqref{eq-Zdef}, $H$ in \eqref{eq-Hdef}, ${H}_y$ in the right after \eqref{eq-ZyZy}, $\Sigma$ in \eqref{eq-SigmaFinal}, and $\Sigma_{C,D}$ in \eqref{eq-SigmaCD}, we define the following notations for the system (\ref{eq-statespace})
\begin{align}
    \hat{Z}&=\begin{bmatrix}
        \hat{A} & \hat{B}\\
        Q^{\frac{1}{2}}\hat{C} & 0\\
        0 & R^{\frac{1}{2}}
        \end{bmatrix},\nonumber\\
    \hat{H}&=[\hat{X}_{+}^\top, (Q^{\frac{1}{2}}\tilde{Y})^\top, (R^{\frac{1}{2}}\tilde{U})^\top, -\hat{X}^\top, -\tilde{U}^\top]^\top,\nonumber\\
    \hat{H}_y&=[\tilde{Y}^\top,  -\hat{X}^\top]^\top,\nonumber\\
    \hat{\Sigma} &=\!\{(\hat{A}, \hat{B}, \hat{C})|[I, \hat{Z}] \hat{N} [I, \hat{Z}]^\top \!= 0\} \text{ with } \hat{N}\!=\!\hat{H}\hat{H}^\top,\nonumber\\
    \hat{\Sigma}_{\hat{C}} &= \{\hat{C}| [I, \hat{C}] \hat{N}_y [I, \hat{C}]^\top \!= 0\} \text{ with } \hat{N}_y=\hat{H}_y\hat{H}_y^\top.\nonumber
\end{align}    

Then, following a similar process in Sections \ref{sc-problemfor}--\ref{sc-mainresults}, we obtain the following two corollaries.
\begin{corollary}\label{co-th3}
    Consider the system (\ref{eq-system}) subject to the input and output constraints (\ref{eq-constraints}). Suppose that Assumption \ref{as-as1} is satisfied and $D=0$. Given a set of pre-collected input-output data (${u}_{[0,T-1]}$, $y_{[0,T-1]}$) of (\ref{eq-system}), if there exist non-negative scalars $\bar{\alpha}_k$ and $\tau_k$, positive scalars $\eta_k$,  $\bar{\beta}_k$ and $\kappa_k$, a matrix $\bar{S}_k\in\mathbb{R}^{m\times\hat{n}}$, and a positive definite matrix $\bar{\Gamma}_k\in\mathbb{R}^{\hat{n}\times\hat{n}}$ at time $k$ such that the following optimization problem (\ref{pr-dpc_con_io}) is feasible,
    \begin{mini!}|s|
        {\substack{\bar{\alpha}_k, \bar{\beta}_k, \eta_k, \tau_k, \\  \kappa_k, \bar{S}_k, \bar{\Gamma}_k}} {\eta_k\label{eq-cor-obj_con}}{\label{pr-dpc_con_io}}{}
        \addConstraint{(\ref{eq-inputconsrx}),}{}
        \addConstraint{\begin{bmatrix}
            -1 & \hat{x}(k|k)^\top\\
            \hat{x}(k|k) & -\bar{\Gamma}_k
            \end{bmatrix}\leq 0,}\label{eq-cor-gammaxcon}{}
            \addConstraint{\bar{\mathcal{M}}_k-\bar{\alpha}_k\hat{\mathcal{N}}\leq \begin{bmatrix}
                -\bar{\beta}_k I_{\hat{n}+m+p} & 0\\
                0 & 0
            \end{bmatrix},}\label{eq-cor-dVwNcon}{}
        \addConstraint{\mathcal{M}_{y,k}-\tau_k\hat{\mathcal{N}}_y\leq \begin{bmatrix}
            -\kappa_k I_{p} & 0\\
            0 & 0
        \end{bmatrix}, }\label{eq-cor-outputconsrx}{}  
    \end{mini!}
    then $\hat{x}(k|k)^\top \eta_k \bar{\Gamma}_k^{-1} \hat{x}(k|k)$ is an upper bound of the objective function $J(k)$ for all systems in $\hat{\Sigma}$, where $\bar{\mathcal{M}}_k$ is defined in (\ref{eq-mathcalMk}), $\mathcal{M}_{y,k}$ is defined in (\ref{eq-mathcalMky}), $\hat{\mathcal{N}}=\text{diag}(\hat{N},0_{\hat{n}\times\hat{n}})$ and $\hat{\mathcal{N}}_y=\text{diag}(\hat{N}_y,0_{\hat{n}\times\hat{n}})$. Further, the state feedback control gain matrix $F_k=\bar{S}_k\bar{\Gamma}_k^{-1}$ minimizes the upper bound. Additionally, the predicted inputs and outputs satisfy the constraints (\ref{eq-inputcons}) and (\ref{eq-outputcons}) for all $\hat{C}$ in $\hat{\Sigma}_{\hat{C}}$ and $i=0,\dots, L-1$.
\end{corollary}
\begin{pf}
    The proof is similar to that of Theorem \ref{th-th3} and thus is omitted.
\end{pf}

\begin{corollary}\label{co-fea_con}
    Consider the system (\ref{eq-system}) subject to the input and output constraints (\ref{eq-constraints}). Suppose that Assumption \ref{as-as1} is satisfied and $D=0$. Given a set of pre-collected input-state-output data (${u}_{[0,T-1]}$, $y_{[0,T-1]}$) of (\ref{eq-system}), if the problem (\ref{pr-dpc_con_io}) is feasible at the initial time $k=n$, then
    \begin{enumerate}[(i)]
        \item the problem (\ref{pr-dpc_con_io}) is feasible at any time $k \geq n$,
        \item the output of the system (\ref{eq-system}) converges to zero under the designed controller $u(k)=F_k\hat{x}(k)$ with $F_k=\bar{S}_k\bar{\Gamma}_k^{-1}$, $k \geq n$. Furthermore, if the system (\ref{eq-system}) is observable, then it is stabilized by the designed controller.
    \end{enumerate}
\end{corollary}
\begin{pf}
    The proof of the statement (i) is similar to that of Theorem \ref{th-fea} and thus is omitted. Following the proof procedure of the statement (ii) of Theorem \ref{th-fea}, we have $\lim_{k\to \infty} \hat{x}(k)=0$. From (\ref{eq-rdef}), we have $\lim_{k\to \infty} u(k)=0$ and $\lim_{k\to \infty} y(k)=0$. Furthermore, when the system (\ref{eq-system}) is observable, we have $\lim_{k\to \infty} x(k)=0$, which means that the designed controller stabilizes (\ref{eq-system}).
\end{pf}

Based on Corollary \ref{co-th3} and Corollary \ref{co-fea_con}, the robust data-driven predictive control scheme using input-output data for the unknown system (\ref{eq-system}) with input and output constraints is summarized in Algorithm \ref{al-dpc_con_io}.

\begin{algorithm}
    \caption{Robust Data-Driven Predictive Control Scheme Using Input-Output Data}\label{al-dpc_con_io}
    \begin{algorithmic}[1]
        \State Collect the input-output sequence;
        \State Set $k=n$;
        \Loop
            \State Set $\hat{x}(k|k)$ with (\ref{eq-rdef});
            \State Solve the problem (\ref{pr-dpc_con_io});
            \State Set $F_k=\bar{S}_k\bar{\Gamma}_k^{-1}$;
            \State Apply the input $u(k|k)=F_k\hat{x}(k|k)$;
            \State Set $k=k+1$;
        \EndLoop
    \end{algorithmic}
\end{algorithm}

\begin{remark}
Notably, there are two main differences between Algorithm \ref{al-dpc_con_io} and Algorithms \ref{al-dpc}-\ref{al-dpc_con}. First, in Algorithm \ref{al-dpc_con_io}, the state $\hat{x}(k|k)$ is constructed using input-output data, and so is the obtained controller. Second, the start time in Algorithm \ref{al-dpc_con_io}  is $k=n$ rather than $k=0$ used in Algorithms \ref{al-dpc}-\ref{al-dpc_con} because the first $n$ input-output data are required to construct the first step of the state $\hat{x}(k|k)$. 
\end{remark}

\begin{remark}
Similar to Remark \ref{re-compareDMPC}, our proposed Algorithm \ref{al-dpc_con_io} offers the advantage of achieving the control target using less data in comparison to the input-output data-based data-driven predictive control schemes proposed in \cite{BerberichJulian2021DMPC, CoulsonJeremy2018DPCI}. Additionally, there is no need to verify the controllability of the controlled plant beforehand; instead, we only need to assess the feasibility of the problem \eqref{pr-dpc_con_io}.
\end{remark}

\section{Case Study}\label{sc-numericalEx}
In this section, we use an unstable batch reactor \cite{WalshGC2001Sonc} to demonstrate the effectiveness of the proposed methods. The discrete-time model (\ref{eq-system}) of the batch reactor has the following parameters with the
  sampling time being set to $0.1s$\cite{Liu2023}
\begin{align}
    &\begin{bmatrix}
        A & B\\
        C & D\\
    \end{bmatrix}\nonumber\\
=&\begin{bmatrix}
   \begin{array}{cccc|cc}
    1.178  &  0.002  &  0.512  & -0.403  &  0.005  & -0.088\\
   -0.052  &  0.662  & -0.011  &  0.061  &  0.467  &  0.001\\
    0.076  &  0.335  &  0.561  &  0.382  &  0.213  & -0.235\\
   -0.001  &  0.335  &  0.089  &  0.849  &  0.213  & -0.016\\
       \hline
       1     & 0     & 1     & -1      & 0 & 0 \\
       0     & 1     & 0     & 0      & 0 & 0 
   \end{array} 
\end{bmatrix}\!\!.\nonumber
\end{align}
These system matrices are presumed to be unknown and used to build the unknown system in the simulation. We assume that the state is unmeasurable; the output measurements are noise-free; and the input and output are subject to constraints: $\|u(k)\|_2\leq \sqrt{2}$ and $\|y(k)\|_2\leq \sqrt{0.2}$, respectively. We choose $Q=10I_2$ and $R=I_2$ in the objective function $J(k)$, and set initial state $x_0=[0.1,0.12,0,-0.1]^\top$ in the simulation. 

Due to the unmeasurable state, we apply the robust data-driven predictive control (RDPC) method developed in Section \ref{sc-arx}, which only uses input-output data. With $n=4$, the past 4-step inputs and outputs before time $k$ are used to construct the state $\hat x(k)$ as
\begin{align}
\hat{x}(k) =[&{u}(k-4)^\top,\dots,{u}(k-1)^\top, \nonumber\\
&y(k-4)^\top, \dots, y(k-1)^\top]^\top.\nonumber
\end{align}
With $m=p=2$ and the definition of $\hat x(k)$, we have  $\hat{n}=16$. Subsequently, a state-space model (\ref{eq-statespace}) with unknown system matrices can be built based on the constructed state. We compare our method with the data-driven predictive control (DPC) scheme, i.e., Algorithm 1 of \cite{BerberichJulian2021DMPC}, by using the same weight matrices in the objective function. 

 To implement our RDPC method, we collect an input-output trajectory (${u}_{[0,17]}$, $y_{[0,17]}$) of the batch reactor with $T=18$, where the input signals ${u}_{[0,17]}$ take values randomly in the interval $[-0.1,0.1]$.  As discussed in Remark \ref{rmrk1}, we set $L=\infty$ for the proposed RDPC scheme. Notably, $T=18$ is the minimum length of the pre-collected data that we can get in this example to make LMIs in Corollary \ref {co-th3} feasible.

According to \cite{BerberichJulian2021DMPC}, the theoretically minimum prediction horizon is $L=4$ and the corresponding minimum length of the pre-collected data for the DPC scheme is $T=35$. However, in our simulation, the DPC is not feasible for $T=35$, and the minimum length of the pre-collected data that we can get through simulation is $T=39$. Further, the control performance of DPC with $T=39$ is not as good as our method with $T=18$. To get a similar control performance as ours, we increase the prediction horizon $L$ to $18$, and the corresponding minimum data length $T$ increases to $81$.

\begin{figure}[htbp]
    \centering
    \begin{subfigure}{0.98\linewidth}
    \centering
    \includegraphics[width=\linewidth]{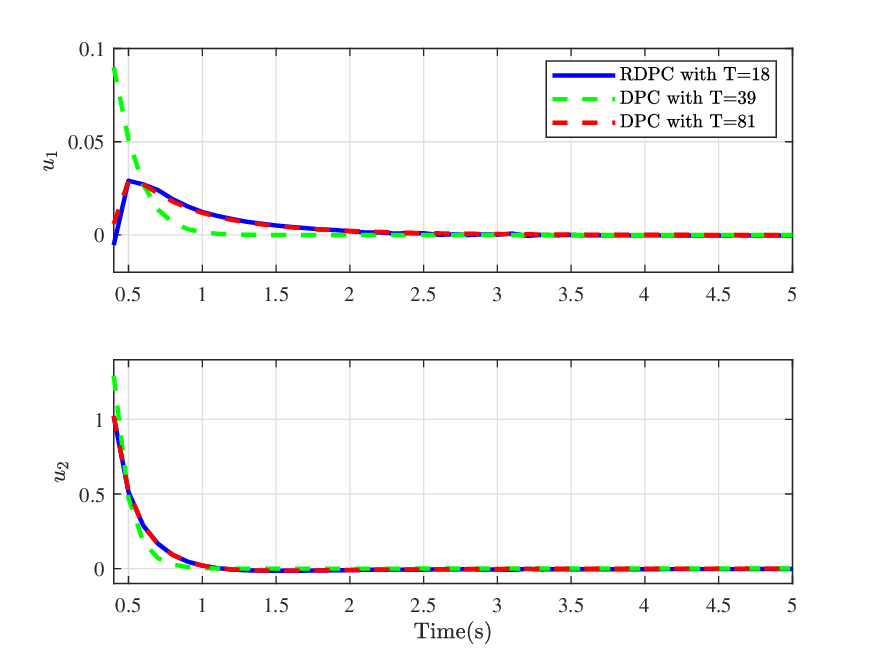}
    \caption{input}
    \end{subfigure}
    \begin{subfigure}{0.98\linewidth}
    \centering
    \includegraphics[width=\linewidth]{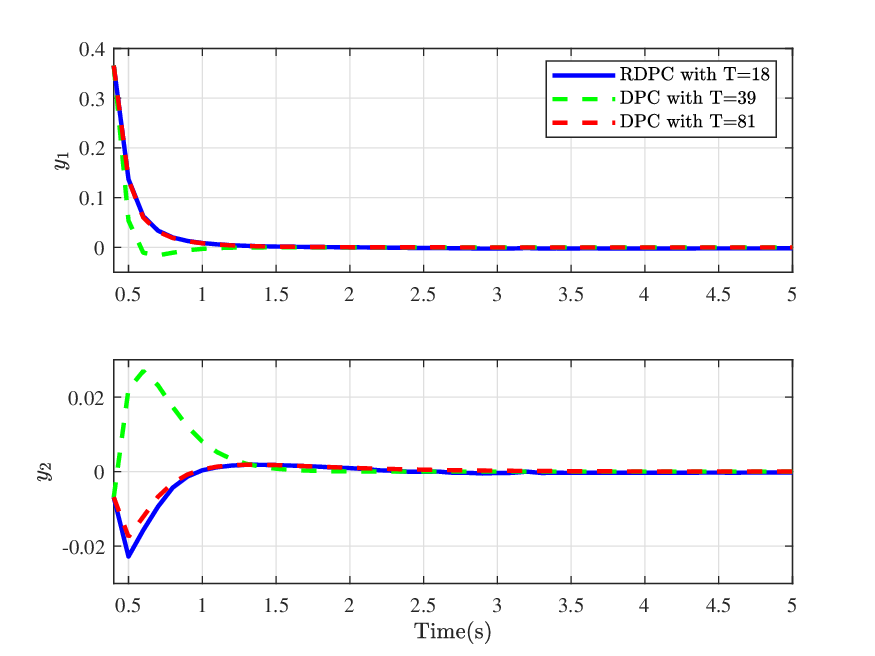}
    \caption{output}
    \end{subfigure}
    \caption{The input and output results of the closed-loop system under the RDPC, and DPC with two different lengths of data.}
    \label{Fig.outputCom}
\end{figure}

The control inputs and the corresponding outputs of the closed-loop system under different control methods are given in Fig. \ref{Fig.outputCom}, where blue lines are the simulation results for the proposed RDPC scheme; green and red dashed lines are the simulation results DPC scheme with lengths of $T=39$ and $T=81$, respectively. It shows that both the control methods achieve the control targets. However, as we already mentioned, the proposed RDPC scheme with $T=18$ has a very similar control performance as that of the DPC scheme with $T=81$ while outperforming the DPC scheme with $T=39$. 

Additionally, when using random inputs generated within the interval $[-0.1,0.1]$ in the simulation, it is hard to collect long enough input-output data. This is because a longer random input increases the possibility of making the output of the unstable batch reactor violate the constraint or even diverge, e.g., when $T\geq 28$ with some given inputs in the simulation, the system output will diverge. To collect a sufficiently long input-output trajectory for the DPC scheme and exclude the potential impact of different data on the control performance, we use the previously collected (${u}_{[0,17]}$, $y_{[0,17]}$) for the first 18 steps data and then apply our proposed RDPC scheme that stabilizes the batch reactor to generate the rest input signal for DPC scheme.  This ensures that the first $18$ input-output pairs of the pre-collected data for the DPC scheme are the same as those for the RDPC scheme. Further, to make the input signal satisfy the PE condition, we also add a random noise within the range of $[-0.1,0.1]$ to the input generated by the RDPC scheme. 

The total costs during the simulation time, i.e., $\bar{J}=\sum_{k=4}^{50}y(k)^\top Q y(k)+u(k)^\top R u(k)$ are given in Table \ref{table1}. 
Similar to the results shown in Fig. \ref{Fig.outputCom}, our method has a very close cost as that of DPC with $T=81$, but has much less cost than DPC with $T=39$. These highlight that our proposed method can achieve comparable control performance with a reduced data length compared to the DPC scheme.

\begin{table}[htbp]
\caption{The objective function values for the RDPC, and DPC with two different lengths of data.}
\centering
\label{table1}
\begin{tabular}{cccccc}
\toprule
Methods & \makecell[c]{RDPC with\\ $T=18$} & \makecell[c]{DPC with \\ $T=39$} & \makecell[c]{DPC with \\ $T=81$}\\
\midrule
$\bar{J}$ & 3.9123 & 4.2257 & 3.9103\\
\bottomrule
\end{tabular}
\end{table}

\section{Conclusion}\label{sc-conclusion}
This paper has investigated the robust data-driven predictive control problem for unknown LTI systems by using input-state-output data and input-output data, depending on whether the system states are measurable. Compared with the existing data-driven predictive control scheme, the proposed scheme does not require the data to satisfy the PE condition of a sufficiently high order. The effectiveness of the proposed methods has been demonstrated through an unstable batch reactor. 

\bibliographystyle{plain}
\bibliography{hukaijian}

\begin{thebibliography}{10}

\bibitem{BerberichJulian2021DMPC}
J.~Berberich, J.~Köhler, M.~A. Müller, and F.~Allgöwer.
\newblock Data-driven model predictive control with stability and robustness
  guarantees.
\newblock {\em IEEE Trans. Automatic Control}, 66(4):1702--1717, 2021.

\bibitem{BerberichJulian2021LtMf}
J.~Berberich, J.~Köhler, M.~A. Müller, and F.~Allgöwer.
\newblock Linear tracking {MPC} for nonlinear systems—part ii: The
  data-driven case.
\newblock {\em IEEE Trans. Automatic Control}, 67(9):4406--4421, 2022.

\bibitem{BisoffiAndrea2020Dsou}
A.~Bisoffi, C.~De Persis, and P.~Tesi.
\newblock Data-based stabilization of unknown bilinear systems with guaranteed
  basin of attraction.
\newblock {\em Systems \& Control Letters}, 145:104788, 2020.

\bibitem{Bolognani2015}
S.~Bolognani, R.~Carli, G.~Cavraro, and S.~Zampieri.
\newblock Distributed reactive power feedback control for voltage regulation
  and loss minimization.
\newblock {\em IEEE Trans. Automatic Control}, 60(4):966--981, 2015.

\bibitem{1994Lmii}
S.~Boyd, L.~E. Ghaoui, E.~Feron, and V.~Balakrishnan.
\newblock {\em Linear Matrix Inequalities in System and Control Theory}.
\newblock Society for Industrial and Applied Mathematics, 1994.

\bibitem{BruntonStevenL2019DSaE}
S.~L. Brunton and J.~N. Kutz.
\newblock {\em Data-Driven Science and Engineering: Machine Learning, Dynamical
  Systems, and Control}.
\newblock Cambridge University Press, New York, 2019.

\bibitem{CamachoE.F2004Mpc}
E.~F. Camacho and C.~Bordons.
\newblock {\em {Model Predictive Control}}.
\newblock Springer, London;, 2nd ed. edition, 2004.

\bibitem{ChenYangquan1999Ilc}
Y.~Chen.
\newblock {\em Iterative Learning Control: Convergence, Robustness, and
  Applications}.
\newblock Springer, London; New York, 1999.

\bibitem{CoulsonJeremy2018DPCI}
J.~Coulson, J.~Lygeros, and F.~Dörfler.
\newblock Data-enabled predictive control: In the shallows of the {DeePC}.
\newblock In {\em Proc. of the European Control Conference}, pages 307--312,
  2019.

\bibitem{Essick2014}
R.~Essick, J.~Lee, and G.~E. Dullerud.
\newblock Control of linear switched systems with receding horizon modal
  information.
\newblock {\em IEEE Trans. Automatic Control}, 59(9):2340--2352, 2014.

\bibitem{GoodwinGrahamC2009AFPa}
G.~C. Goodwin and K.~S. Sin.
\newblock {\em Adaptive Filtering Prediction and Control}.
\newblock Dover Publications, New York, 2009.

\bibitem{HouZhongsheng2014MFAC}
Z.~Hou and S.~Jin.
\newblock {\em Model Free Adaptive Control: Theory and Applications}.
\newblock CRC Press, Bosa Roca, 2014.

\bibitem{hu2022}
K.~Hu and T.~Liu.
\newblock Data-driven {$H_\infty$} control for unknown linear time-invariant
  systems with bounded disturbances.
\newblock In {\em Proc. of the IEEE Conference on Decision and Control}, pages
  1423--1428, 2022.

\bibitem{Hu2023}
K.~Hu and T.~Liu.
\newblock Data-driven output-feedback control for unknown switched linear
  systems.
\newblock {\em IEEE Control Systems Letters}, 7:2299--2304, 2023.

\bibitem{He2021}
H.~Kong, M.~Shan, S.~Sukkarieh, T.~Chen, and W.~X. Zheng.
\newblock Kalman filtering under unknown inputs and norm constraints.
\newblock {\em Automatica}, 133:109871, 2021.

\bibitem{KothareMayureshV1996Rcmp}
M.~V. Kothare, V.~Balakrishnan, and M.~Morari.
\newblock Robust constrained model predictive control using linear matrix
  inequalities.
\newblock {\em Automatica}, 32(10):1361--1379, 1996.

\bibitem{Liu2023}
W.~Liu, J.~Sun, G.~Wang, F.~Bullo, and J.~Chen.
\newblock Data-driven resilient predictive control under denial-of-service.
\newblock {\em IEEE Trans. Automatic Control}, 68(8):4722--4737, 2023.

\bibitem{Lopez2023}
V.~G. Lopez, M.~Alsalti, and M.~A. Müller.
\newblock Efficient off-policy {Q}-learning for data-based discrete-time {LQR}
  problems.
\newblock {\em IEEE Trans. Automatic Control}, 68(5):2922--2933, 2023.

\bibitem{LuppiAlessandro2022Odso}
A.~Luppi, C.~De~Persis, and P.~Tesi.
\newblock On data-driven stabilization of systems with nonlinearities
  satisfying quadratic constraints.
\newblock {\em Systems \& Control Letters}, 163:105206, 2022.

\bibitem{MAYNE2000789}
D.~Q. Mayne, J.~B. Rawlings, C.~V. Rao, and P.~O.~M. Scokaert.
\newblock Constrained model predictive control: Stability and optimality.
\newblock {\em Automatica}, 36(6):789--814, 2000.

\bibitem{DePersisClaudio2020FfDC}
C.~De Persis and P.~Tesi.
\newblock Formulas for data-driven control: Stabilization, optimality, and
  robustness.
\newblock {\em IEEE Trans. Automatic Control}, 65(3):909--924, 2020.

\bibitem{DePersisClaudio2021LloL}
C.~De Persis and P.~Tesi.
\newblock Low-complexity learning of linear quadratic regulators from noisy
  data.
\newblock {\em Automatica}, 128:109548, 2021.

\bibitem{PhanMP1996Rbsa}
M.~P. Phan and R.~W. Longman.
\newblock Relationship between state-space and input-output models via observer
  markov parameters.
\newblock {\em WIT Trans. the Built Environment}, 19, 1996.

\bibitem{QIN2003733}
S.~J. Qin and T.~A. Badgwell.
\newblock A survey of industrial model predictive control technology.
\newblock {\em Control Engineering Practice}, 11(7):733--764, 2003.

\bibitem{RakoviSasaV2018HoMP}
S.~V. Raković and W.~S. Levine.
\newblock {\em Handbook of Model Predictive Control}.
\newblock Springer Basel AG, Cham, 2018.

\bibitem{SteentjesTomRV2022Odci}
T.~R.~V. Steentjes, M.~Lazar, and P.~M.~J. Van~den Hof.
\newblock On data-driven control: Informativity of noisy input-output data with
  cross-covariance bounds.
\newblock {\em IEEE Control Systems Letters}, 6:2192--2197, 2022.

\bibitem{Talebi2022}
S.~Talebi, S.~Alemzadeh, N.~Rahimi, and M.~Mesbahi.
\newblock On regularizability and its application to online control of unstable
  {LTI} systems.
\newblock {\em IEEE Trans. Automatic Control}, 67(12):6413--6428, 2022.

\bibitem{TangiralaArunK2015PoSI}
A.~K. Tangirala.
\newblock {\em Principles of System Identification: Theory and Practice}.
\newblock CRC Press, 2015.

\bibitem{Waarde2023}
H.~J. van Waarde, M.~K. Camlibel, J.~Eising, and H.~L. Trentelman.
\newblock Quadratic matrix inequalities with applications to data-based
  control.
\newblock {\em SIAM Journal on Control and Optimization}, 61(4):2251--2281,
  2023.

\bibitem{vanWaardeHenkJ2020Fndt}
H.~J. van Waarde, M.~K. Camlibel, and M.~Mesbahi.
\newblock From noisy data to feedback controllers: Nonconservative design via a
  matrix s-lemma.
\newblock {\em IEEE Trans. Automatic Control}, 67(1):162--175, 2022.

\bibitem{vanWaardeHenkJ2020WFLf}
H.~J. van Waarde, C.~De Persis, M.~K. Camlibel, and P.~Tesi.
\newblock Willems’ fundamental lemma for state-space systems and its
  extension to multiple datasets.
\newblock {\em IEEE Control Systems Letters}, 4(3):602--607, 2020.

\bibitem{WalshGC2001Sonc}
G.~C. Walsh and Hong Y.
\newblock Scheduling of networked control systems.
\newblock {\em IEEE Control Systems Magazine}, 21(1):57--65, 2001.

\bibitem{WillemsJanC2005Anop}
J.~C. Willems, P.~Rapisarda, I.~Markovsky, and B.~L. M.~De Moor.
\newblock A note on persistency of excitation.
\newblock {\em Systems \& Control Letters}, 54(4):325--329, 2005.

\end{thebibliography}

\end{document}